\documentclass{article}

\usepackage{cite}
\usepackage{amsmath,amssymb,amsfonts}
\usepackage{caption}
\usepackage{algorithmic}
\usepackage{graphicx}
\usepackage{textcomp}
\usepackage{booktabs}
\usepackage{mathrsfs}
\usepackage{xcolor}
\usepackage{multirow}
\usepackage{indentfirst}
\usepackage{subfigure}
\usepackage{colortbl}
\definecolor{mygray}{gray}{.9}
\usepackage{threeparttable}
\usepackage{color, soul}
\usepackage{float}
\usepackage{tabularx}
\usepackage{xspace}
\usepackage{enumitem}
\usepackage{makecell}
\usepackage{stfloats}
\usepackage{placeins}
\usepackage{xurl}
\usepackage[normalem]{ulem}
\definecolor{darkpink}{HTML}{D1477F}
\definecolor{darkgreen}{RGB}{0,130,0}

\newcommand{\sys}[0]{MIAO}

\usepackage{hyperref}

\usepackage[accepted]{mlsys2025}
\makeatletter
\renewcommand{\mlsys@appearing}{}
\makeatother
\hypersetup{pdfsubject={}}

\mlsystitlerunning{MIAO: Microscaling Intermediate Accumulation Optimization}

\begin{document}

\twocolumn[
\mlsystitle{MIAO: Microscaling Intermediate Accumulation Optimization}

\mlsyssetsymbol{equal}{*}

\begin{mlsysauthorlist}
\mlsysauthor{Chloe Wong}{equal,ucb}
\mlsysauthor{Nicolas Rakela}{equal,ucb}
\mlsysauthor{Man Shi}{ucb}
\mlsysauthor{Hansung Kim}{ucb}
\mlsysauthor{Ruohan Richard Yan}{ucb}
\mlsysauthor{Coleman Richard Charles Hooper}{ucb}
\mlsysauthor{Yakun Sophia Shao}{ucb}
\end{mlsysauthorlist}

\mlsysaffiliation{ucb}{University of California, Berkeley, Berkeley, CA, USA}

\mlsyscorrespondingauthor{Chloe Wong}{chloe.wong@berkeley.edu}
\mlsyscorrespondingauthor{Nicolas Rakela}{nicorakela@berkeley.edu}

\mlsyskeywords{Microscaling, Quantization, Accumulator precision, Systolic arrays, Hardware accelerators, LLM inference}
\vskip 0.3in

\begin{abstract}

Microscaling (MX) quantization enables lower-precision large language model (LLM) inference with limited accuracy degradation, reducing memory requirement and multiplication cost.
However, despite reduced operand precision, accumulation is often retained at high floating-point precisions, leaving a substantial hardware cost that operand quantization alone does not address.
In particular, systolic arrays, as used in tensor accelerators such as Google TPU and AWS Trainium, distribute accumulation across distinct processing elements but conventionally use uniform accumulator precision throughout the array. 
This spatial organization offers an opportunity to reduce hardware cost by tailoring precision to individual accumulation stages. 

In this work, we present \sys{}, a hardware-software co-design framework that jointly selects product precision and stage-dependent accumulator widths to minimize hardware area and energy cost under an end-to-end perplexity constraint.
MIAO first calibrates MX block-scale shifts to reduce intermediate magnitudes while preserving operand quantization accuracy, then assigns narrower accumulators to earlier stages and progressively wider accumulators to later stages. 
To efficiently explore the resulting design space, \sys{} combines hardware cost estimates derived from logic synthesis with a calibrated accuracy predictor to identify promising precision configurations, then validates the accuracy of promising candidates through bit-exact end-to-end LLM evaluation.
Across MXFP8, MXFP6, and MXFP4, \sys{} reduces processing element (PE) array area by up to 38.7\% and energy by up to 36.1\% relative to the all-BF16 accumulation baseline, with at most 1.60\% additional perplexity degradation over operand-only MX quantization across six LLMs. 

\end{abstract}
]

\printAffiliationsAndNotice{\mlsysEqualContribution} %
{\let\thefootnote\relax\footnotetext{Distribution Statement A: Approved for public release. Distribution is unlimited.}}

\section{Introduction}
\label{intro}

\begin{figure*}[t]
  \centering
  \includegraphics[width=\textwidth]{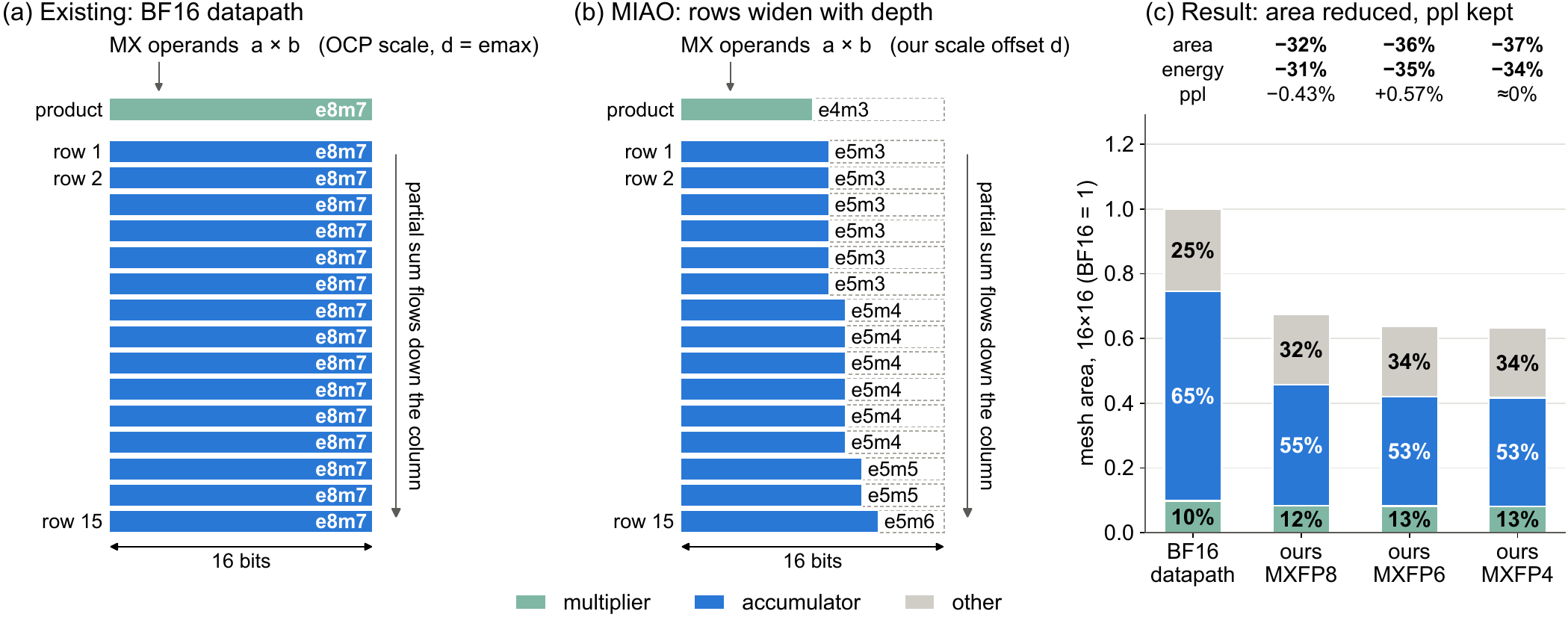}
  \vspace{-20pt}
  \caption{\textbf{MIAO reduces intermediate arithmetic precision by narrowing accumulator widths row by row, cutting PE-array area by 32-37\% and energy by 31-35\% while preserving end-to-end accuracy.}
  (a) Conventional BF16 intermediate arithmetic.
  (b)~MIAO: calibrated MX scaling, reduced product precision, row-dependent accumulation (MXFP4 example).
  (c)~Area breakdown from our area model. Perplexity: TinyLlama-1.1B on WikiText-2, vs.\ OCP-scaled MX; energy: whole accelerator, per operation.}

  \label{fig:headline}
  \vspace{-14pt}
\end{figure*}

Large language models (LLMs) impose substantial computational and memory demands, motivating the development of specialized accelerators for efficient inference \citep{jouppi2017datacenter}. A key technique for improving hardware efficiency is quantization, which reduces the precision of weights and activations to lower memory requirements and allows less costly arithmetic operations \citep{horowitz2014computing}. 
Aggressive low-bit quantization can introduce substantial numerical error due to the wide dynamic ranges of weights and activations, particularly in the presence of outliers \citep{xiao2023smoothquant,dettmers2022llmint8}.
Microscaling (MX) formats mitigate this error by sharing a scale across small blocks of low-precision floating-point elements, adapting their representation to local value ranges and enabling aggressive operand quantization with limited accuracy degradation~\citep{ocp2023mx}.

However, simply reducing operand precision does not eliminate the cost of intermediate arithmetic. Products and partial sums often retain wider floating-point precision to accommodate their dynamic range and limit rounding error.
In particular, floating-point accumulation requires exponent alignment, significand arithmetic, and normalization, making it a substantial component of the datapath even when the operands use only a few bits. 
In our evaluated baseline designs, accumulation accounts for more than 50\% of PE area and up to 54\% of PE energy.
These costs limit the hardware savings achievable through operand quantization alone \citep{abdelaziz2021rethinking}.

Prior work has addressed low-precision inference through operand scaling, accumulation-aware quantization, and intermediate arithmetic hardware optimizations.
VS-Quant improves quantization accuracy through vector-wise scaling~\citep{dai2021vsquant}, and \citet{chhugani2026mxfp4} similarly improve MXFP4 operand scaling without optimizing intermediate arithmetic precision.
Accumulator-aware quantization methods such as A2Q, A2Q+, and AXE adapt operand quantization to satisfy predefined accumulator constraints
\citep{colbert2023a2q,colbert2024a2qplus, colbert2025axe}.
Other work directly reduces floating-point arithmetic overhead
by optimizing product and alignment widths~\citep{abdelaziz2021rethinking} or
exploring precision bounds and hierarchical accumulation~\citep{he2026precision}.
However, these approaches do not explicitly optimize floating-point accumulator width across individual stages of an MX systolic array.

Our key insight is that accumulation precision can be tailored to individual hardware stages rather than chosen uniformly for the entire array.
Conventional designs nevertheless use the same accumulator format throughout the array. 
For example, in weight-stationary systolic arrays, as used in tensor accelerators such as Google's TPU and AWS's Trainium, each processing element multiplies an incoming activation by a stationary weight and adds the product to a partial sum received from the preceding row~\cite{kwon2019understanding,sze2017efficient}.
Successive accumulation stages therefore map to distinct hardware units whose widths can be selected independently. 
Earlier stages accumulate fewer products, creating an opportunity to use narrower formats, while later stages can retain wider formats to accommodate growing partial sums and limit rounding error \cite{higham2002accuracy}.
We exploit this spatial organization to reduce hardware cost while preserving end-to-end model accuracy.

Realizing this opportunity, however, requires more than just independently narrowing each accumulator; naively reducing intermediate arithmetic precision in MX accelerators can lead to substantial accuracy degradation.
First, intermediate precision depends on MX operand scaling.
Conventional MX block scaling places quantized operands near the upper range of their operand format to limit underflow, producing large products and partial sums that require wide floating-point accumulation to avoid overflow.
Reducing the block-scale factor can lower these magnitudes but may increase operand quantization error.
Second, rounding errors introduced at different accumulation stages propagate through subsequent computations and affect end-to-end model accuracy.
As a result, product and accumulation precisions must be selected jointly, creating a combinatorial design space that is expensive to explore through exhaustive LLM evaluation \cite{dong2019hawq}.

To address these limitations, we present \sys{}, a hardware-software co-design framework that jointly selects product precision and stage-dependent accumulator widths to minimize hardware area and energy cost under end-to-end perplexity constraints. 
\sys{} first calibrates MX block-scale shifts to reduce intermediate product and partial-sum magnitudes while preserving operand quantization accuracy.
\sys{} then combines hardware cost estimates derived from logic synthesis with a calibrated accuracy predictor to search over configurations with narrower accumulation in earlier stages and progressively wider accumulators in later stages.
The predictor guides the search toward promising configurations, while bit-exact end-to-end LLM evaluation verifies whether each candidate meets the perplexity constraint. 

Our contributions are:
\begin{enumerate}[nosep, leftmargin=*]
    \item  We propose stage-dependent floating-point accumulation for MX systolic arrays, replacing uniform accumulator precision with widths tailored to individual stages of the reduction. Together with block-scale calibration and product precision selection, this enables narrower arithmetic hardware while preserving end-to-end model accuracy.
    
    \item We develop an automated search methodology that combines synthesis-derived hardware costs, an accuracy predictor, and bit-exact LLM evaluation to select precision configurations under a perplexity constraint. For MXFP8, it reduces the search space by \textbf{177× and 3,009× relative to exhaustive enumeration} on 16×16 and 32×32 systolic arrays, respectively.
    
    \item We demonstrate hardware implementations that reduce PE-array \textbf{area by up to 38.7\%} and measured \textbf{energy per operation reduction by up to 36.1\%} relative to an all-BF16 datapath baseline, with \textbf{at most 1.6\% }additional perplexity degradation across six LLMs.

\end{enumerate}
\section{Background}
\label{sec:bg}

\subsection{Microscaling Formats and Arithmetic}
\label{sec:bg:microscaling}

Microscaling (MX) formats represent tensors using low-precision floating-point elements with a shared scale for each small block. Compared to coarser-grained quantization, smaller blocks allow scales to better match local value ranges, reducing quantization error and enabling lower-precision representations of LLM weights and activations. We consider MXFP8 (E4M3), MXFP6 (E3M2), and MXFP4 (E2M1), following the Open Compute Project (OCP) MX specification \cite{ocp2023mx}. 

Each MX block contains 32 elements and a shared 8-bit exponent-only (E8M0) scale, representing powers of two as $x = 2^{E-127}$, where $E$ is the encoded exponent. This enables efficient hardware scaling through exponent adjustment rather than general floating-point multiplication.

For a block of values $\mathbf{v}$, MX quantization computes a shared scale $x$, quantized elements $\mathbf{P}$, and reconstructed values $\bar{\mathbf{v}}$ as
\begin{equation}
    x = 2^{\lfloor \log_2 \max_i |v_i| \rfloor - d},
    \quad
    \mathbf{P} = Q\left(\frac{\mathbf{v}}{x}\right),
    \quad
    \bar{\mathbf{v}} = x\mathbf{P},
    \label{eq:mx_quant}
\end{equation}
where $Q(\cdot)$ rounds to the target element format and $d$ is an integer shift controlling the shared scale. The OCP specification fixes $d=e_{\max}$, the exponent of the largest normal value of the element format, with $e_{\max}=8$, $4$, and $2$ for MXFP8, MXFP6, and MXFP4, respectively.

For two MX blocks $\mathbf{A}$ and $\mathbf{B}$, their dot product is: 
\begin{equation}
    \operatorname{dot}(\mathbf{A},\mathbf{B})
    = x_A x_B \sum_{i=1}^{k} P_i^{(A)}P_i^{(B)},
    \label{eq:mx_dot}
\end{equation}
where $x_A$ and $x_B$ are the corresponding block scales and $k=32$. Longer dot products combine contributions from multiple blocks, each with its own scale factors. While the MX specification defines numerical representation and dot-product semantics, it leaves intermediate product and accumulation precision to the implementation~\cite{ocp2023mx}. Existing implementations commonly retain BF16 or FP32 accumulation despite using low-precision MX operands \citep{wipfli2026vmxdotp,wang2026mfsa}.

In this work, we apply MX quantization to both weights and activations during LLM inference. Each linear layer computes $Y = W \times A$ using MX-quantized operands, and its output is requantized into MX blocks before use by subsequent MX-quantized operations.

\subsection{Floating-Point Accumulation in Systolic Arrays}

\begin{figure}[H]
    \centering
    \includegraphics[width=\linewidth]{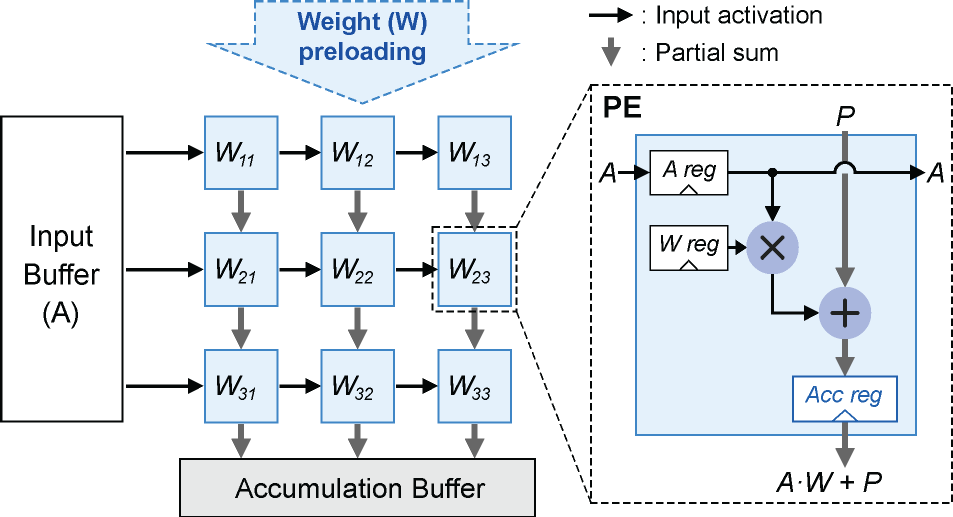}
    \vspace{-20pt}
    \caption{Weight-stationary systolic array dataflow, PE datapath. 
    }
    \vspace{-8pt}
    \label{fig:WS_array}
\end{figure}

Systolic arrays accelerate matrix multiplication using grids of processing elements (PEs) performing parallel multiply-accumulate (MAC) operations. Common dataflows include output-stationary (OS), where each PE accumulates a local output, and weight-stationary (WS), where weights remain fixed in PEs while activations propagate horizontally and partial sums propagate vertically \citep{chen2016eyeriss}. We focus on WS arrays (Figure~\ref{fig:WS_array}), where successive accumulation stages map to different PE rows. This allows each row to use a fixed accumulator precision, enabling stage-dependent precision allocation through differently sized arithmetic hardware.

In the WS dataflow, each PE multiplies an incoming activation by its stationary weight and adds the product to the partial sum received from the preceding row. While low-precision operands reduce multiplier costs, accumulators typically retain wider floating-point formats to accommodate the growing range of partial sums and limit rounding error. Floating-point accumulation requires exponent alignment and significand arithmetic, contributing substantially to hardware cost \citep{mach2021fpnew}. As shown in Fig.~\ref{fig:headline}(c), BF16 accumulation accounts for approximately 65\% of PE-array area in our baseline with MX operands and BF16 intermediate arithmetic.

Because each successive row adds another product, partial-sum magnitudes generally increase with accumulation depth. This suggests that different stages may have different precision requirements, motivating the characterization of stage-dependent accumulation precision in Sec.~3.

\section{Datapath Product and Accumulation Precision Optimization}
\label{method}

\begin{figure*}
  \centering
  \includegraphics[width=\textwidth]{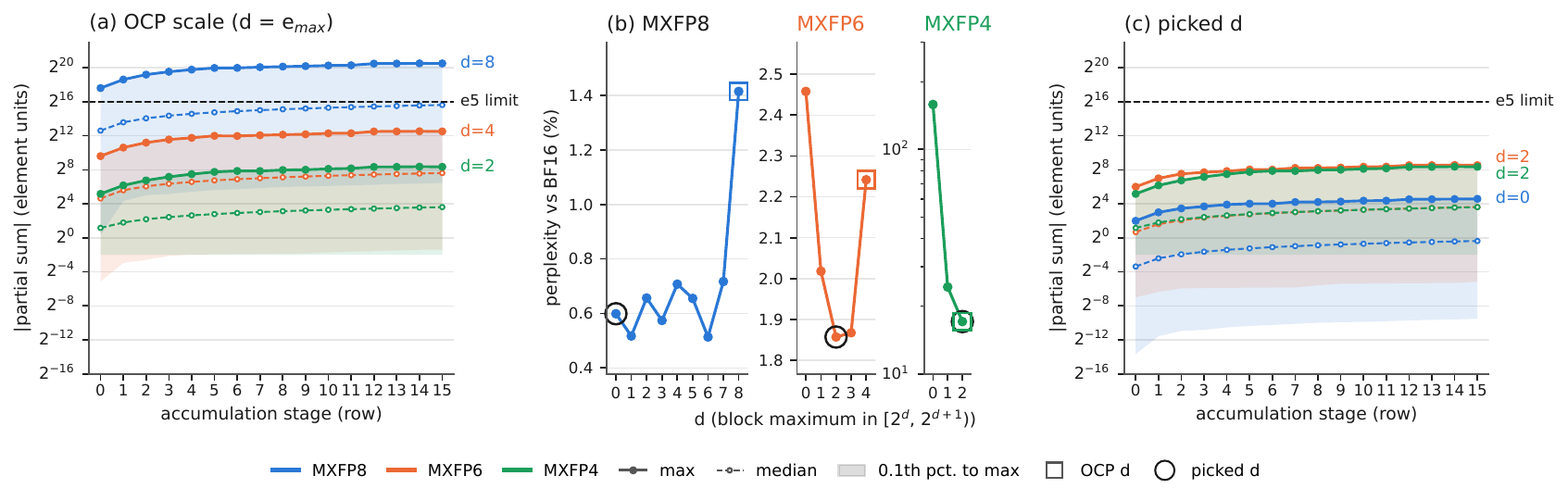}
  \vspace{-27pt}
  \caption{Choosing the block-scale shift $d$ (TinyLlama, one 16-row column).
    (a)~and~(c) show partial sums at each accumulation stage, before and after choosing $d$.
    (b)~The picked $d$ is the smallest $d$ whose perplexity is within noise of the best.}
   \vspace{-10pt}
  \label{fig:d_story}
\end{figure*}

We first examine why conventional OCP block-scaling limits opportunities for intermediate precision reduction (Sec.~3.1), then show how calibrating the prescribed block-scale shifts can reduce precision requirements (Sec.~3.2). We then use a weight-stationary systolic array as a case study to investigate how reduced product precision and row-dependent floating-point accumulation can translate into hardware area and energy savings (Sec. 3.3). These observations motivate the automated precision optimization method in Sec.~4.

\subsection{Limitations of Existing Microscaling}

The intermediate precision requirements of MX arithmetic depend not only on the operand format, but also on the block-scale shift $d$ (Eq. ~\ref{eq:mx_quant}). The OCP specification sets $d=e_{\max}$, with $e_{\max}=8$, $4$, and $2$ for MXFP8, MXFP6, and MXFP4, respectively~\citep{ocp2023mx}. This places the largest scaled element near the upper range of its format to limit underflow. However, because OCP's scaling rule is defined independently of intermediate arithmetic precision, it can impose substantial precision requirements on the hardware datapath.

\textbf{Large intermediate dynamic range.}
Increasing $d$ by one doubles the magnitude of each scaled operand and increases product magnitudes by a factor of four, before rounding and saturation. For MXFP8, the OCP shift $d=8$ therefore increases nominal product magnitudes by a factor of $2^{16}$ relative to $d=0$. This amplification carries into intermediate partial sums, which must be represented before the block scales are applied to rescale the dot-product result.

Figure~\ref{fig:d_story}(a) shows partial-sum magnitudes under OCP scaling across a 16-row weight-stationary systolic array. For TinyLlama, partial sums reach approximately $2^{20}$, $2^{12}$, and $2^8$ for MXFP8, MXFP6, and MXFP4, respectively. These magnitudes impose substantial exponent-range requirements on intermediate accumulation, limiting opportunities to narrow accumulator formats without risking overflow.

\textbf{Operand quantization error.}
The OCP scaling rule can also introduce saturation error. Setting $d=e_{\max}$ places the block maximum in the highest power-of-two range of the element format, which is not fully representable. Consequently, scaled elements exceeding the maximum representable value are clipped. For TinyLlama under MXFP8, the maximum element is clipped in 14-16\% of activation blocks and 9-13\% of weight blocks, with an average clipping magnitude of 8\%. This error persists even when intermediate arithmetic uses high precision.

Together, these limitations make conventional OCP scaling poorly suited to reducing intermediate arithmetic precision: large partial-sum magnitudes restrict accumulator narrowing, while operand saturation introduces accuracy loss even with high-precision arithmetic. Exploiting narrower product and accumulator formats therefore requires reconsidering the OCP-defined block-scale shift $d$.
 
\subsection{MX Block-Scale Shift Calibration}

To address the limitations of conventional OCP scaling, we calibrate the block-scale shift $d$ separately for each MX format. From Eq.~\ref{eq:mx_quant}, the scaled block maximum satisfies
\begin{equation}
    2^d \leq \frac{\max_i |v_i|}{x} < 2^{d+1}.
    \label{eq:mx_normalized_range}
\end{equation}
before element rounding and saturation. Reducing $d$ decreases product and partial-sum magnitudes, enabling narrower intermediate arithmetic, but can increase operand underflow. We therefore balance operand quantization accuracy against intermediate precision requirements.

\textbf{Calibration Procedure.}
For each MX format, we sweep integer shifts $d \in \{0,\ldots,d_{\mathrm{OCP}}\}$, where $d_{\mathrm{OCP}}$ is the OCP-defined shift. We measure end-to-end perplexity using MX-quantized operands with full-precision matrix multiplication, isolating operand quantization error from reduced intermediate precision. We select the smallest $d$ whose perplexity is statistically indistinguishable from the minimum measured perplexity, reducing product and downstream partial-sum magnitudes while preserving quantization accuracy.

Fig.~\ref{fig:d_story}(b) shows the perplexity across block-scale shifts for TinyLlama, from which we select $d=0$, $2$, and $2$ for MXFP8, MXFP6, and MXFP4, respectively. Comparing Fig.~\ref{fig:d_story}(a,c), these calibrated shifts substantially reduce partial-sum magnitudes relative to OCP scaling for MXFP8 and MXFP6, while MXFP4 retains its OCP-defined shift. The selected shifts remain fixed during subsequent precision optimization. Nevertheless, partial-sum magnitudes still generally grow with accumulation depth, indicating that precision requirements vary across rows. This motivates configuring accumulator precision independently for each row rather than using a uniform format throughout the array. 

\begin{figure*}[t]
    \centering
    \hspace{-1.1em}
    \includegraphics[width=1.02\textwidth]{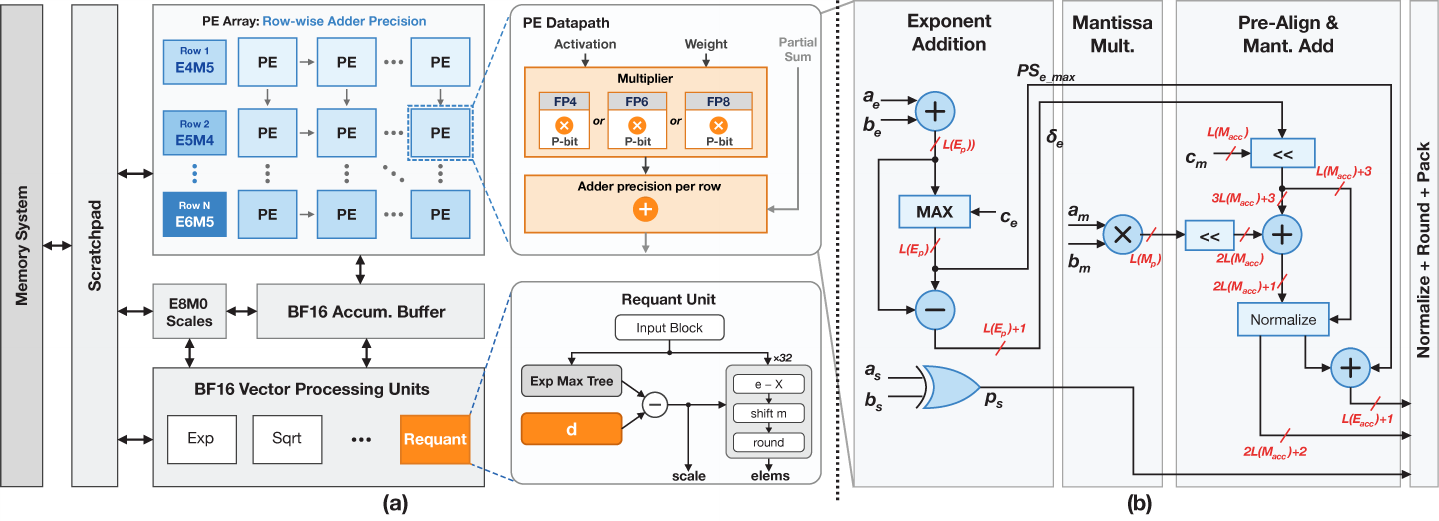}
    \vspace{-20pt}
    \caption{
    \textbf{Proposed Hardware Architecture.}
    \textbf{(a)} Precision-configurable systolic array datapath with three design parameters: \textbf{(1)} MX block-scale shift $d$, calibrated per operand format; \textbf{(2)} product floating-point precision, shared across PEs; and \textbf{(3)} row-wise accumulator floating-point precision, independently configurable at each accumulation stage. Inter-tile accumulation retains BF16 precision.
    \textbf{(b)} Processing element microarchitecture implementing a fused multiply-add (FMA) operation. $L(\cdot)$ denotes configurable wire bit-widths.
    }
    \vspace{-14pt}
    \label{fig:full-datapath}
\end{figure*}

\subsection{Precision-Configurable Systolic Datapath and Hardware Characterization}

To investigate the hardware benefits of the precision reduction opportunities identified in Sec. 3.2, we parameterize a weight-stationary systolic array with configurable floating-point formats for products and row-wise accumulation. This serves as a hardware case study for characterizing how intermediate arithmetic precision affects area and energy.

\vspace{\baselineskip}

\textbf{Precision-Configurable Datapath.}

We extend the open-source Gemmini accelerator~\citep{genc2021gemmini} to support MX operands and configurable intermediate floating-point precision in its weight-stationary systolic array (Fig.~\ref{fig:full-datapath}). Specifically, we support two configurable precision parameters:

\textbf{1. Configurable product precision.}
Each PE multiplies an MX-quantized weight and activation, rounding its product to configurable floating-point format $f_p$, which is shared across all PEs. 

\textbf{2. Row-wise accumulation precision.} For an $n$-row array, the $n-1$ successive accumulation stages use independently configurable formats $f_1,\ldots,f_{n-1}$. At stage $i$, the incoming partial sum is added to the product, and the result is rounded to format $f_i$. This allows accumulator precision to vary across rows. 

\textbf{3. MX requantization and scale handling.}
At each tile's output, the accumulated result is rescaled using the corresponding MX block scales (Eq ~\ref{eq:mx_dot}). Rescaled outputs from multiple tiles are accumulated in BF16. Once the full matrix multiplication is complete, the result is requantized to the target MX format before being used by subsequent MX-quantized operations.

\textbf{Hardware Precision Characterization.}

Now, we characterize the numerical requirements of intermediate products and partial sums and quantify the hardware costs of reduced floating-point precision.

\textbf{Product profiling.}

For each MX format, we derive the minimum floating-point format required to represent all quantized element products without rounding. With the calibrated block-scale shifts, these formats are E5M7, E4M5, and E4M3 for MXFP8, MXFP6, and MXFP4, respectively. These establish reference points for exploring narrower product formats that will introduce rounding error.

\textbf{Accumulator profiling.}

We capture quantized operands from 67 linear layers across 16 evaluation samples and simulate systolic accumulation using exact products and high-precision partial sums. We record partial-sum magnitudes and exponent distributions at each stage. For MXFP8 at $d=0$, the maximum observed magnitude grows from 8 after the first addition to 23.93 after the fifteenth, suggesting there are different exponent-range requirements across accumulation stages.

\textbf{Hardware sensitivity.}
To quantify the cost of accumulation precision, we independently vary exponent and mantissa widths, applying each format uniformly across the 15 accumulation stages of a $16\times16$ systolic array. Table~\ref{tab:per-bit} reports the resulting changes in PE-array area, energy, and perplexity for MXFP8.

\begin{table}[!ht]
\centering\small
\setlength{\tabcolsep}{4pt}
\begin{tabular}{lrrrr}
\toprule
Lane width & \shortstack{PE-array area\\(k$\mu$m$^2$)} & \shortstack{PE-array \\area} & Energy & \shortstack{Perplexity\\(\% points)} \\
\midrule
\multicolumn{5}{l}{\emph{mantissa, lanes e5, relative to m7}} \\
m7 & 0 & 0.0\% & 0.0\% & +0.00 \\
m6 & $-$46.8 & $-$6.7\% & $-$8.2\% & +0.03 \\
m5 & $-$88.4 & $-$12.7\% & $-$13.9\% & +0.05 \\
m4 & $-$114.1 & $-$16.4\% & $-$19.2\% & +0.70 \\
m3 & $-$155.8 & $-$22.4\% & $-$24.5\% & +2.20 \\
m2 & -- & -- & -- & +8.42 \\
\midrule
\multicolumn{5}{l}{\emph{exponent, lanes m7, relative to e8}} \\
e8 & 0 & 0.0\% & 0.0\% & +0.00 \\
e7 & $-$7.8 & $-$1.1\% & $-$0.9\% & +0.00 \\
e6 & $-$17.7 & $-$2.5\% & $-$1.6\% & +0.00 \\
e5 & $-$18.4 & $-$2.6\% & $-$0.9\% & +0.00 \\
e4 & $-$20.2 & $-$2.9\% & $-$3.1\% & +0.07 \\
\bottomrule
\end{tabular}
\caption{\textbf{Accumulator exponent and mantissa sensitivity for MXFP8.} Hardware and perplexity changes are relative to the widest format in each sweep (e5m7 for the mantissa sweep; e8m7, i.e.\ BF16, for the exponent sweep). Hardware costs are evaluated using the synthesis-calibrated model described in Sec.~4.2.}
\label{tab:per-bit}
\end{table}

Reducing accumulator mantissa width from M7 to M5 decreases PE-array area by 12.7\% and energy by 13.9\%, compared with 2.6\% and 0.9\% reductions from narrowing the exponent width from E8 to E5. Thus, reducing mantissa width offers substantially greater hardware savings.

These characterizations establish the numerical requirements and hardware costs used to select product and row-wise accumulator formats under end-to-end accuracy constraints in Sec.~4.

\section{Framework and Search}

Building on the precision-configurable datapath and hardware characterization in Sec.~3, we develop an automated method for selecting product and accumulation precisions under end-to-end accuracy constraints. Sec.~4.1 formulates the optimization problem, Sec.~4.2 describes the hardware-software evaluation method, and Sec.~4.3 presents the automated search, loss-model prediction, and certification procedure.

\subsection{Problem Formulation}

We formulate product and accumulation precision selection as a constrained hardware cost minimization problem. For an $n\times n$ weight-stationary systolic array, we define the precision configuration
\begin{equation}
    \theta = (f_p, f_{a,1}, \ldots, f_{a,n-1}),
    \label{eq:precision_config}
\end{equation}
where $f_p$ is the product format shared across PEs and $f_{a,r}$ is the accumulator format at row $r$. The accumulator formats form a non-decreasing precision ladder along each reduction path. We consider MXFP8, MXFP6, and MXFP4 operands with their respective calibrated block-scale shifts from Sec.~3.2.

For each configuration $\theta$, we evaluate end-to-end perplexity $\mathrm{PPL}(\theta)$, PE area $A(\theta)$, and energy per operation $E(\theta)$. We minimize the normalized area-energy product:

\begin{equation}
    H(\theta) =
    \frac{A(\theta)}{A_{\mathrm{BF16}}}
    \cdot
    \frac{E(\theta)}{E_{\mathrm{BF16}}}.
    \label{eq:hardware_cost}
\end{equation}

where $A_{\mathrm{BF16}}$ and $E_{\mathrm{BF16}}$ are the area and energy costs of a baseline datapath that uses the same MX-quantized operands and array dimensions, but performs all product and accumulation operations in BF16.

We seek the minimum-cost configuration satisfying a specified accuracy constraint:
\begin{equation}
\begin{aligned}
    \underset{\theta \in \Theta}{\text{minimize}}
    \quad & H(\theta) \\
    \text{subject to} \quad &
    \mathrm{PPL}(\theta)
    \leq (1+\epsilon)\mathrm{PPL}_{\mathrm{BF16}}.
\end{aligned}
\label{eq:optimization}
\end{equation}

Here, $\Theta$ denotes the supported precision configurations, $\mathrm{PPL}_{\mathrm{BF16}}$ is the perplexity of the unquantized BF16 reference model, and $\epsilon$ specifies the maximum allowable relative perplexity increase. This constraint accounts for degradation from both MX operand quantization and reduced intermediate precision. Varying $\epsilon$ allows us to explore the trade-off between hardware cost and end-to-end accuracy.

Although the formulation generalizes to other accumulation topologies, including adder trees, we focus on weight-stationary systolic arrays in this work (Sec.~3.3).

\subsection{Hardware-Software Evaluation Framework}

Evaluating each precision configuration requires end-to-end LLM inference and hardware cost characterization, making exhaustive exploration expensive. To reduce evaluation overhead, we combine bit-exact arithmetic simulation, accelerated LLM inference, and synthesis-derived hardware area and energy modeling.

\textbf{Bit-Exact Arithmetic Simulation.} 
We implement a configurable arithmetic simulator that reproduces the numerical behavior of the systolic datapath in Sec.~3.3. The simulator models MX operand quantization, product rounding to the selected format $f_p$, row-wise accumulation using the specified formats $f_{a,r}$, and BF16 inter-tile accumulation. Operations follow the datapath's order of execution, including intermediate rounding, saturation, and underflow behavior. We validate the simulator against hardware arithmetic outputs, obtaining bit-identical results across 65,536 evaluated elements for MXFP8.

\textbf{End-to-end LLM integration.} We integrate the simulator into LLM inference by replacing selected linear-layer matrix multiplications with configurable MX arithmetic while retaining BF16 for other operations. This enables direct measurement of perplexity for each precision configuration. An optimized Triton implementation preserves the simulator's bit-exact outputs and accelerates repeated end-to-end perplexity evaluations by 2.2--7.4$\times$ on TinyLlama.

\textbf{Hardware area and energy estimation.} We construct a hardware area and energy model from logic synthesis and gate-level power analysis of parameterized processing elements (PEs) in a 16\,nm technology. We sweep product formats with exponent widths of 3--8 bits and significand widths of 3--8 bits, and accumulator formats with exponent widths of 4--8 bits and significand widths of 4--12 bits. Using these synthesis results, the model combines PE costs according to each configuration's product precision and row-wise accumulator formats, enabling efficient area and energy estimation across precision configurations and array sizes without synthesizing each candidate.

\subsection{Automated Precision Search and Certification}
\label{sec:precision_search}

We develop an automated framework (Algorithm~\ref{alg:precision_search}) that takes a target perplexity constraint and selects a low-cost product and accumulator precision configuration without manual tuning. 

\textbf{1. Search Space.}
We first constrain the search space to keep exploration tractable. For an $n\times n$ weight-stationary array, we impose two restrictions:\\
(i) We use a shared product format and a non-decreasing ladder of $n-1$ accumulator formats, reflecting increasing partial-sum magnitudes and rounding sensitivity across rows. \\
(ii) We fix accumulator exponent widths to five bits, sufficient for the observed dynamic range, and vary mantissa widths from three to seven bits to capture the accuracy-hardware cost trade-off (Sec.~3.3).\\
These restrictions yield $\binom{n+3}{4}$ ladders: 330, 3,876, and 52,360 for $8\times8$, $16\times16$, and $32\times32$ arrays, respectively. For $32\times32$ certification, we additionally restrict precision transitions to four-row boundaries to reduce search complexity, yielding 495 ladders per product format. Two-step ladders from candidate selection (step 3) keep any transition row.

\textbf{2. Feasibility Screening.}
To avoid unnecessary precision search, we first check whether the selected MX operand format can meet the perplexity target. We evaluate the calibrated operands with lossless products and BF16 accumulation, terminating the search if measured perplexity exceeds the target beyond evaluation noise.

\textbf{3. Candidate Selection.}
To establish an initial low-cost feasible configuration and collect data for loss-model fitting, we fix product format and sweep uniform accumulator mantissa widths to identify two adjacent widths that bracket the perplexity target: the narrowest width that meets the target and the next narrower width that exceeds it. Using these two widths, we construct two-step ladders by assigning the narrower precision to rows $1,\ldots,k$ and the wider precision to rows $k+1,\ldots,n-1$. We sweep the transition row $k$ to identify the lowest-cost feasible ladder.

\textbf{4. Loss-Model-Guided Refinement.}
To efficiently explore multi-step ladders, we fit a model that predicts perplexity for untested ladders using measurements from the initial sweeps. The model assumes that rounding the partial sum $S_r$ to $m_r$ bits introduces zero-mean noise with variance proportional to $\mathbb{E}[S_r^2]4^{-m_r}$~\cite{Widrow96}. Assuming uncorrelated rounding errors, their variances add across rows; a second-order loss approximation relates these variances to expected loss increase~\cite{Dong20}. We model the resulting log-perplexity increase as 
\begin{equation}
    \log\frac{\widehat{\mathrm{PPL}}(\theta)}
              {\mathrm{PPL}_{\mathrm{base}}}
    =
    \sum_{r=1}^{n-1}
    w_r\left(4^{-m_r}-4^{-7}\right),
    \label{eq:accuracy_prediction}
\end{equation}
where $w_r$ is the fitted coefficient for rounding at row $r$, and $\mathrm{PPL}_{\mathrm{base}}$ is the measured perplexity for the same product format with BF16 accumulation.

To fit the model, we approximate $w_r\propto\mathbb{E}[S_r^2]\propto r+1$, assuming approximately constant loss sensitivity across rows. Partial-sum profiling follows this trend within 9\% across 67 layers. We compare a single-scale fit to the measured row profile against non-negative, smoothness-regularized per-row coefficients, selecting by five-fold cross-validation. On $8\times8$ MXFP8, fitting 11 ladders predicts 319 unseen ladders with an RMS perplexity error of 0.0066, below the paired-run standard error of 0.0085.\\

\begin{algorithm}[t]
\caption{Automated Precision Search}
\label{alg:precision_search}
\begin{algorithmic}[1]
\REQUIRE Calibrated MX format $f_{\mathrm{MX}}$, array size $n$, \\perplexity tolerance $\epsilon$, noise margin $\delta$
\ENSURE Best feasible configuration $\theta^*$, certification status
\STATE $T \gets (1+\epsilon)\mathrm{PPL}_{\mathrm{BF16}}$

\STATE \textbf{[1] Search Space:} Construct non-decreasing precision ladders

\STATE \textbf{[2] Feasibility Screening:}
\IF{operand-only PPL with exact products and BF16 accumulation $>T+\delta$}
    \STATE \textbf{return} infeasible
\ENDIF

\STATE \textbf{[3] Candidate Selection:}
\FOR{each product format $f_p$}
    \STATE Find adjacent uniform accumulator widths bracketing $T$
    \STATE Sweep two-step transitions; update $\theta^*$
\ENDFOR

\STATE \textbf{[4] Loss-Model-Guided Refinement:}
\STATE Fit loss model to measured configurations
\WHILE{promising lower-cost candidates remain}
    \STATE Predict perplexity and cost; select candidates
    \STATE Evaluate end-to-end; update model and $\theta^*$
\ENDWHILE

\STATE \textbf{[5] Search-Space Certification:}
\STATE Prune cheaper candidates no more precise than measured failures
\STATE Evaluate remaining cheaper candidates end-to-end
\STATE Certify if all cheaper candidates are excluded
\STATE \textbf{return} $\theta^*$, certification status
\end{algorithmic}
\end{algorithm}
\vspace{-20pt}

We first use the fitted loss model to predict perplexity and the hardware cost model (Sec.~4.2) to estimate area and energy for untested ladders. Based on these predictions, we select promising low-cost candidates for bit-exact end-to-end LLM evaluation. We then use the measured perplexity to update the loss model and determine whether each candidate satisfies the accuracy constraint.

\textbf{5. Search-Space Certification.}
Finally, we assess whether the best configuration is optimal within the restricted search space. We identify cheaper candidates and prune those no more precise than previously failed configurations, assuming accuracy does not worsen with increased precision. We evaluate the remaining candidates end-to-end and update the best configuration if a cheaper feasible candidate is found. We certify optimality only when all cheaper candidates are excluded.

Together, these stages provide an automated procedure for selecting product and accumulator precisions under a user-specified perplexity constraint, with explicit verification of the resulting configuration.

\section{Evaluation}
\label{eval}

We evaluate our framework's ability to reduce hardware cost while preserving end-to-end LLM accuracy. Our selected configurations reduce PE-array compute area by 32.0-38.7\% and estimated whole-accelerator energy per operation by 23.7-36.1\% relative to all-BF16, with at most 1.06\% additional perplexity degradation over operand-only MX quantization.

We first describe the experimental setup, then examine hardware-accuracy trade-offs and selected precision configurations. Finally, we evaluate cross-model generalization and downstream task accuracy.

\subsection{Experimental Setup}

\textbf{Software.}
We evaluate TinyLlama-1.1B and five additional LLMs on WikiText-2, using 16 windows of 2,048 tokens per model. We optimize precision configurations on TinyLlama and apply the selected configurations to the remaining models without retuning. Perplexity is measured through end-to-end inference with bit-exact emulation of MX operands, products, and accumulation.

\textbf{Hardware.}
We evaluate $8\times8$, $16\times16$, and $32\times32$ weight-stationary systolic arrays. Hardware costs are estimated using the synthesis-derived model in Sec.~4.2, targeting a 16\,nm technology and a 2\,ns clock period. We additionally use FireSim FPGA-accelerated simulation~\citep{firesim-isca18} to compare hardware arithmetic outputs against the bit-exact software reference.

\textbf{Baselines and Metrics.}
We use two accuracy baselines: (1) an unquantized BF16 model with BF16 products, accumulation, and inter-tile additions, and (2) operand-only MX quantization with BF16 products, accumulation, and inter-tile additions. We report perplexity relative to both baselines, along with PE-array area, estimated whole-accelerator energy per operation, and power. Hardware costs are compared against BF16 intermediate arithmetic with the same MX operands and array dimensions.
 
\subsection{Hardware-Accuracy Trade-offs}
\label{eval:tradeoffs}

\begin{figure*}[!t]
  \centering
  \includegraphics[width=\textwidth]{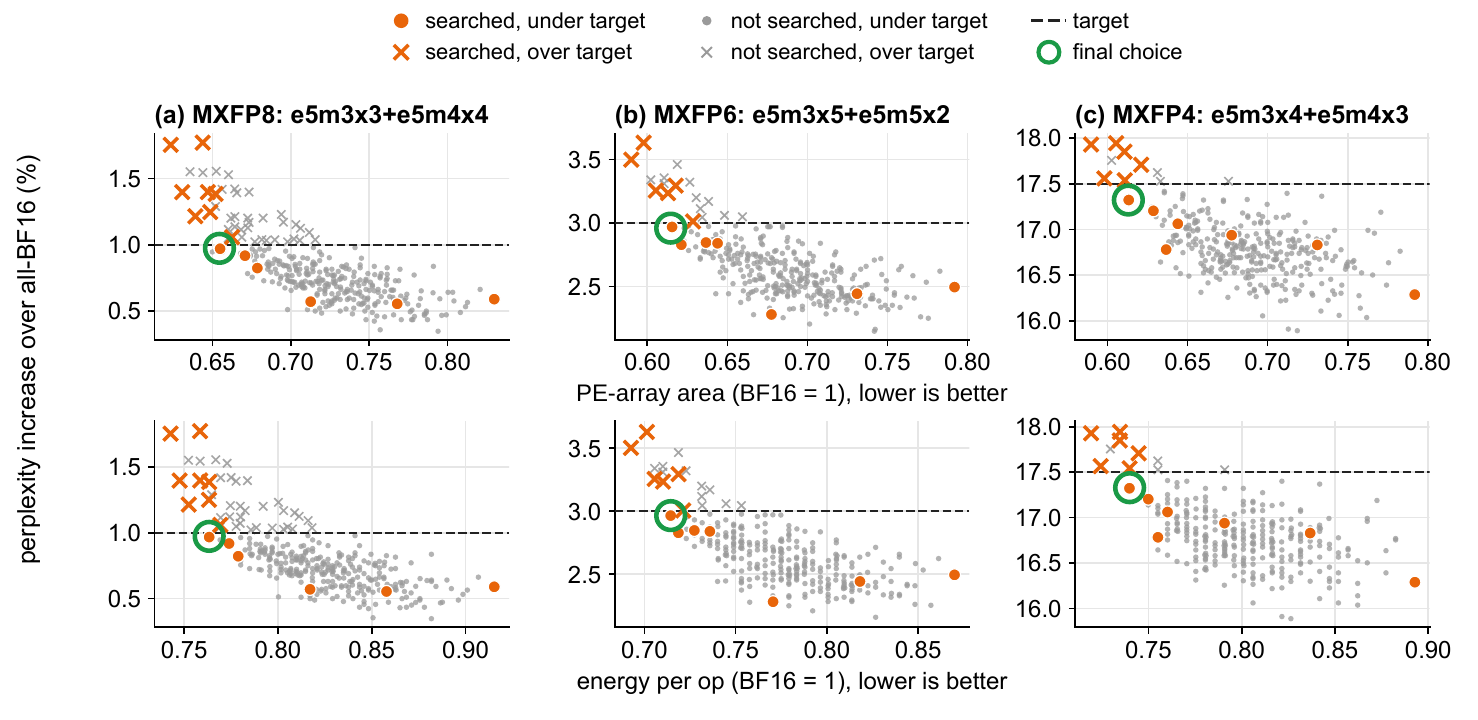}
  \vspace{-30pt}
  \caption{\textbf{Hardware-accuracy trade-offs at $8\times8$.} Orange points are configurations evaluated during search; grey points are additional exhaustive evaluations for verification. Each point represents an accumulator ladder at the selected $d$ and product format. Top: PE-array area; bottom: whole-accelerator energy per operation, both normalized to all-BF16.}
  \vspace{-10pt}
  \label{fig:area-energy-8x8}
\end{figure*}

\textbf{Search Outcomes.}
Figure~\ref{fig:area-energy-8x8} compares the automated search with exhaustive evaluation of the restricted $8\times8$ precision space. The search evaluates only a subset of configurations, while the exhaustive results verify that the selected configurations minimize hardware cost under the perplexity constraint.

\textbf{Selected Precisions.}
Table~\ref{tab:picks} reports the selected product and accumulator formats across MX formats and array sizes. All configurations use progressively wider accumulator mantissas, while product precision and accumulator transitions vary with MX format and array size. 

\begin{table}[!t]
\centering\small
\setlength{\tabcolsep}{3pt}
\newcommand{\p}{\,{+}\,}
\resizebox{\columnwidth}{!}{%
\begin{tabular}{llcll}
\toprule
Format & Size & $d$ & Prod & Accum (rows $1..n{-}1$) \\
\midrule
\multirow{3}{*}{MXFP8}
 & $8{\times}8$ & 0 & e4m6 & e5m3$\times$3\p e5m4$\times$4 \\
 & $16{\times}16$ & 0 & e4m5 & e5m4$\times$11\p e5m5$\times$4 \\
 & $32{\times}32$ & 0 & e4m6 & e5m4$\times$19\p e5m5$\times$12 \\
\midrule
\multirow{3}{*}{MXFP6}
 & $8{\times}8$ & 2 & e4m3 & e5m3$\times$5\p e5m5$\times$2 \\
 & $16{\times}16$ & 2 & e4m4 & e5m3$\times$6\p e5m4$\times$8\p e5m5$\times$1 \\
 & $32{\times}32$ & 2 & e4m4 & e5m3$\times$2\p e5m4$\times$29 \\
\midrule
\multirow{3}{*}{MXFP4}
 & $8{\times}8$ & 2 & e4m3 & e5m3$\times$4\p e5m4$\times$3 \\
 & $16{\times}16$ & 2 & e4m3 & e5m3$\times$6\p e5m4$\times$6\p e5m5$\times$2\p e5m6$\times$1 \\
 & $32{\times}32$ & 2 & e4m3 & e5m3$\times$4\p e5m4$\times$24\p e5m6$\times$3 \\
\bottomrule
\end{tabular}%
}
\caption{\textbf{Selected precision configurations.} $d$: block-scale shift; Prod: product format; Accum: row-wise accumulator formats.}
\label{tab:picks}
\vspace{-15pt}
\end{table}

\textbf{Hardware Cost and Accuracy.}
Table~\ref{tab:results} reports hardware savings and perplexity changes relative to both BF16 and operand-only MX quantization.

(i) MXFP8 reduces PE-array area by \textbf{32.0--34.5\%} and energy per operation by \textbf{23.7--32.5\%}, with at most 0.50\% additional perplexity degradation over operand-only MX. \\
(ii) MXFP6 reduces area by \textbf{34.6--38.5\%} and energy by \textbf{28.6--36.1\%}, with at most 1.06\% additional perplexity degradation over operand-only MX. \\ 
(iii) MXFP4 reduces PE-array area by \textbf{35.2--38.7\%} and energy by \textbf{26.0--36.1\%}, with at most 0.89\% additional perplexity degradation over operand-only MX, despite greater total degradation from operand quantization.

\begin{table*}[!t]
\centering\small
\setlength{\tabcolsep}{4pt}
\begin{tabular}{llccccccccc}
\toprule
Format & Size & Perplexity & \shortstack{$\Delta$ vs\\op.-only} & \shortstack{$\Delta$ vs\\BF16} & \shortstack{Energy\\(pJ/op)} & $\Delta$ & \shortstack{Area\\(mm$^2$)} & $\Delta$ & \shortstack{Power\\(mW)} & $\Delta$ \\
\midrule
\multirow{3}{*}{MXFP8}
 & $8{\times}8$ & 7.3581 & +0.44\% & +0.97\% & 6.39 & -23.7\% & 0.115 & -34.5\% & 197.4 & -23.7\% \\
 & $16{\times}16$ & 7.3036 & +0.38\% & +0.98\% & 4.19 & -30.7\% & 0.471 & -32.4\% & 517.2 & -30.7\% \\
 & $32{\times}32$ & 7.2929 & +0.50\% & +0.97\% & 3.65 & -32.5\% & 1.892 & -32.0\% & 1804.5 & -32.5\% \\
\midrule
\multirow{3}{*}{MXFP6}
 & $8{\times}8$ & 7.5035 & +1.06\% & +2.97\% & 1.65 & -28.6\% & 0.108 & -38.5\% & 203.8$^*$ & -28.6\% \\
 & $16{\times}16$ & 7.4468 & +1.02\% & +2.96\% & 1.12 & -34.5\% & 0.445 & -36.2\% & 553.4$^*$ & -34.5\% \\
 & $32{\times}32$ & 7.4387 & +0.96\% & +2.99\% & 0.99 & -36.1\% & 1.820 & -34.6\% & 1960.0 & -36.1\% \\
\midrule
\multirow{3}{*}{MXFP4}
 & $8{\times}8$ & 8.5495 & +0.89\% & +17.32\% & 1.45 & -26.0\% & 0.108 & -38.7\% & 178.6 & -26.0\% \\
 & $16{\times}16$ & 8.4695 & 0.00\% & +17.10\% & 0.91 & -34.1\% & 0.441 & -36.8\% & 449.6$^*$ & -34.1\% \\
 & $32{\times}32$ & 8.4861 & +0.78\% & +17.50\% & 0.78 & -36.1\% & 1.803 & -35.2\% & 1541.5$^*$ & -36.1\% \\
\bottomrule
\end{tabular}
\caption{\textbf{Hardware--accuracy results on TinyLlama-1.1B.} Perplexity changes are relative to operand-only MX and all-BF16; hardware changes are relative to all-BF16 at the same PE-array size. Area: post-synthesis PE-array area. Energy and power: whole-accelerator estimates. $^*$: power derived from energy per operation.}
\label{tab:results}
\end{table*}

\subsection{Cross-Model Generalization}
\label{eval:generalization}

We apply the $16\times16$ configurations selected on TinyLlama to five additional LLMs without retuning. Table~\ref{tab:cross-model-results} shows that intermediate precision reduction introduces \textbf{at most a 1.6\% relative perplexity} increase over operand-only MX quantization. MXFP4 incurs substantially greater total perplexity degradation relative to BF16, primarily from operand quantization. These results demonstrate transferability across models, although the TinyLlama accuracy constraints are not guaranteed to hold.

\begin{table}[H]
\centering\small
\setlength{\tabcolsep}{4pt}
\begin{tabular}{llccc}
\toprule
Model & Format & Perplexity & \shortstack{$\Delta$ vs\\BF16} & \shortstack{$\Delta$ vs\\op.-only} \\
\midrule
\multirow{4}{*}{Mistral-7B}
 & BF16 & 5.73 & -- & -- \\
 & MXFP8 & 5.76 & +0.5\% & +0.2\% \\
 & MXFP6 & 5.85 & +2.1\% & +0.7\% \\
 & MXFP4 & 6.31 & +10.1\% & +0.5\% \\
\midrule
\multirow{4}{*}{Llama-3.1-8B}
 & BF16 & 6.52 & -- & -- \\
 & MXFP8 & 6.63 & +1.7\% & +0.6\% \\
 & MXFP6 & 6.86 & +5.2\% & +1.5\% \\
 & MXFP4 & 8.36 & +28.3\% & +1.1\% \\
\midrule
\multirow{4}{*}{Llama-2-7B}
 & BF16 & 5.15 & -- & -- \\
 & MXFP8 & 5.18 & +0.6\% & +0.3\% \\
 & MXFP6 & 5.32 & +3.3\% & +1.6\% \\
 & MXFP4 & 5.93 & +15.3\% & +0.5\% \\
\midrule
\multirow{4}{*}{Llama-2-13B}
 & BF16 & 4.66 & -- & -- \\
 & MXFP8 & 4.68 & +0.5\% & +0.2\% \\
 & MXFP6 & 4.74 & +1.7\% & +0.6\% \\
 & MXFP4 & 5.02 & +7.7\% & +0.8\% \\
\midrule
\multirow{4}{*}{Qwen2.5-14B}
 & BF16 & 5.34 & -- & -- \\
 & MXFP8 & 5.44 & +1.8\% & +0.6\% \\
 & MXFP6 & 5.64 & +5.5\% & +1.1\% \\
 & MXFP4 & 6.88 & +28.8\% & +0.5\% \\
\bottomrule
\end{tabular}
\caption{\textbf{Cross-model transfer of TinyLlama-selected $16\times16$ configurations.} WikiText-2 perplexity (16 windows of 2,048 tokens). Changes are relative to all-BF16 and operand-only MX.}
\label{tab:cross-model-results}
\end{table}

\textbf{Downstream Tasks.}
We further evaluate selected configurations on seven zero-shot tasks using TinyLlama and HumanEval code generation using Llama-3.1-8B (Appendix, Tables~\ref{tab:zeroshot} and~\ref{tab:humaneval}). Across all array sizes, MXFP8 and MXFP6 maintain mean zero-shot accuracy within \textbf{0.91 percentage points} of all-BF16. On HumanEval, selected configurations change pass@1 by +3.7, -3.0, and +0.6 percentage points relative to operand-only MXFP8, MXFP6, and MXFP4, respectively, with no statistically significant differences. These results provide further evidence that reduced product and accumulator precision preserves downstream performance beyond perplexity.

\section{Related Work}

\textbf{Mixed-precision allocation.}
  Mixed-precision quantization assigns more bits where the model is most sensitive. HAQ~\citep{Wang19} uses reinforcement learning with hardware feedback to choose layer-wise weight and activation bitwidths, and HAWQ-V2~\citep{Dong20} ranks layers by Hessian trace. For LLMs, AWQ~\citep{lin2024awq} protects salient weight channels through activation-aware scaling, Atom~\citep{zhao2024atom} keeps outlier channels at higher precision within 4-bit serving kernels, SliM-LLM~\citep{huang2025slimllm} allocates bitwidths to weight groups by salience, MixLLM~\citep{zheng2026mixllm} assigns 4- or 8-bit precision to individual output features, and FGMP~\citep{hooper2025fgmp} selects FP8 or NVFP4 per weight and activation block using Fisher-weighted sensitivity, with matching datapath support. All of these allocate precision to operands. \sys{} instead allocates precision to intermediate arithmetic: the product and each accumulator row of a systolic array, at fixed operand formats.

  \textbf{Precision search.}
  Because the space of per-layer precision assignments grows exponentially, recent work reduces search cost. FLIQS~\citep{dotzel2024fliqs} runs a one-shot reinforcement-learning search over integer and floating-point formats, AMQ~\citep{lee2025amq} prunes the search space and guides an iterative search with a learned quality predictor, and KVTuner~\citep{li2025kvtuner} prunes per-layer key and value precision pairs offline using attention-output error. \sys{} shares the goal of avoiding exhaustive evaluation, but its search space is ordered by construction: a fitted rounding-loss model proposes accumulator ladders, and every selected configuration is certified by bit-exact end-to-end evaluation.

  \textbf{Hardware cost models.}
  Analytical models such as Timeloop~\citep{parashar2019timeloop} and Accelergy~\citep{wu2019accelergy} estimate accelerator energy and area from architectural actions and per-component costs, enabling fast design-space exploration. \sys{} prices each configuration with a compositional area and energy model calibrated on post-synthesis processing elements, so that the cost of one additional mantissa or exponent bit in a single accumulator row is resolved.
\section{Conclusion}

We present \sys, a hardware-software co-design framework for jointly optimizing product and accumulator precision in MX-based systolic arrays. By modeling row-dependent rounding sensitivity and hardware cost, \sys{} selects and certifies non-uniform accumulation ladders under a target perplexity constraint. Across MXFP8, MXFP6, and MXFP4, the selected configurations reduce PE-array area by \textbf{32.0-38.7\%} and estimated whole-accelerator energy per operation by \textbf{23.7-36.1\%} relative to all-BF16, with \textbf{at most 1.06\% }additional perplexity degradation beyond operand-only quantization. Configurations selected on TinyLlama transfer to five additional LLMs without retuning, adding at most \textbf{1.6\% perplexity} beyond operand-only quantization. Future work includes cross-model calibration and extending the optimization to other accelerator architectures and workloads.

\section{Acknowledgements}
This research was supported in part by the NSF CAREER Award CCF-2238346
and in part by the Specialized Computing Ecosystems (SLICE) Lab industrial
sponsors and affiliates. This project is also funded by the Microelectronics
Commons Program, a DoW initiative. Distribution Statement A: Approved for
public release. Distribution is unlimited. The views and conclusions
contained in this document are those of the authors and should not be
interpreted as representing the official policies, either expressed or
implied, of the U.S. Government.

\bibliography{example_paper}
\bibliographystyle{mlsys2025}

\appendix

\clearpage

\section{Additional Evaluation Results}
\label{app:additional-evaluation}

\subsection{Area--Accuracy Trade-offs}
\label{app:area-ppl}

\begin{figure*}[!tp]
\centering
\includegraphics[width=0.96\textwidth]{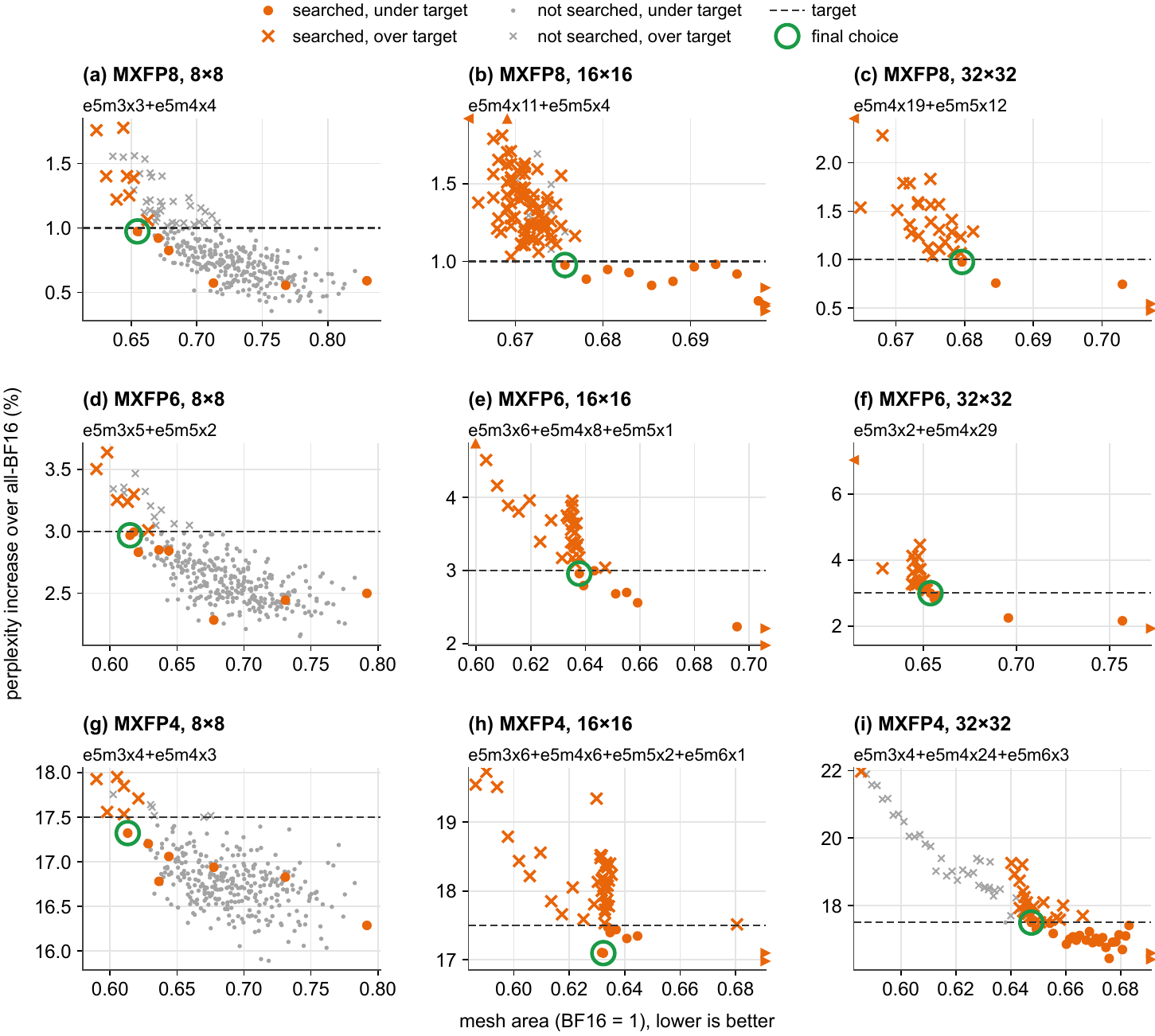}
\caption{\textbf{PE-array area versus perplexity, all formats and array sizes.} Each point is one measured accumulator ladder at the selected $d$ and product format; area is normalized to the all-BF16 PE-array of the same size. Grey ladders were measured outside the search, mainly an exhaustive $8{\times}8$ sweep that validates the search's choice. The $16{\times}16$ and $32{\times}32$ axes are zoomed; triangles mark points beyond the edge.}
\label{fig:grid-area}
\label{fig:grid-area-ppl}
\end{figure*}

Figure~\ref{fig:grid-area-ppl} extends the main-text $8\times8$ analysis to all MX formats and array sizes, showing the measured area--accuracy trade-offs and configurations selected by the search.

\subsection{Energy--Accuracy Trade-offs}
\label{app:energy-ppl}

\begin{figure*}[!tp]
\centering
\includegraphics[width=0.96\textwidth]{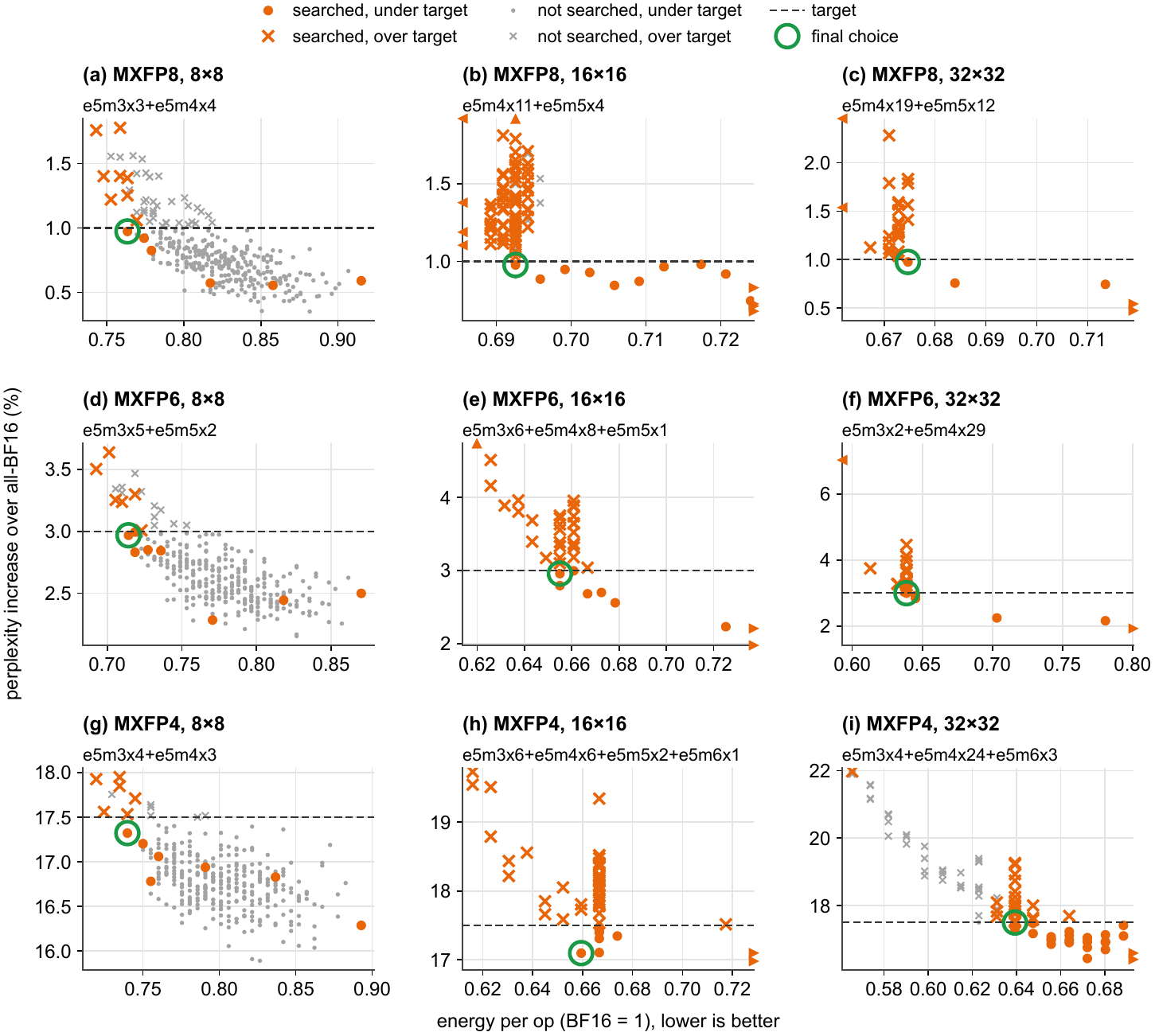}
\caption{\textbf{Energy versus perplexity, all formats and array sizes.} Each point is one measured accumulator ladder at the selected $d$ and product format; energy is whole-accelerator energy per operation, normalized to the all-BF16 datapath. Grey ladders were measured outside the search, mainly an exhaustive $8{\times}8$ sweep that validates the search's choice. The $16{\times}16$ and $32{\times}32$ axes are zoomed; triangles mark points beyond the edge.}
\label{fig:grid-energy}
\label{fig:grid-energy-ppl}
\end{figure*}

Figure~\ref{fig:grid-energy-ppl} presents the corresponding energy--accuracy trade-offs across all nine configurations.

\subsection{Cross-Model Perplexity and Performance}
Tables~\ref{tab:results-mistral-7b}--\ref{tab:results-qwen25-14b} apply TinyLlama-selected configurations (Table~\ref{tab:picks}) to five LLMs without retuning or re-imposing perplexity targets. Perplexity uses WikiText-2 (16 windows of 2,048 tokens, seed 0), with MLP linears and \texttt{lm\_head} mapped to the array. $\Delta$ vs op.-only compares against MX operands with lossless products and BF16 accumulation; all other $\Delta$ values compare against all-BF16 at the same size. Hardware area, energy, and power match Table~\ref{tab:results}; $^*$ marks power derived from energy per operation.

FireSim cycle-accurate simulation~\cite{firesim-isca18} measures a 32-token prefill after a 2,048-token KV cache, with BF16 attention. Cycle $\Delta$ compares against a BF16 systolic array (negative is faster); reported cycles equal BF16 cycles times $(1+\Delta)$.

\subsection{Zero-Shot Task Accuracy}
\label{app:zeroshot}

Table~\ref{tab:zeroshot} evaluates seven tasks on TinyLlama-1.1B. Selected MXFP8 and MXFP6 configurations change mean accuracy by at most 0.91 points relative to all-BF16; MXFP4 loses 2.31, 2.03 and 2.46 points at $8\times8$, $16\times16$ and $32\times32$. Across all nine configurations, selected product and accumulator precisions change mean accuracy by at most 0.7 points relative to operand-only MX. 

\subsection{Code Generation}

Table~\ref{tab:humaneval} tests whether the selected configurations hold up on generation, where errors can compound over many decoding steps. We run Llama-3.1-8B (base) on all 164 HumanEval problems (lm-eval, 0-shot, greedy decoding, pass@1) at $16{\times}16$, with the MLP linears and lm\_head through the array and attention in BF16, as in the cross-model runs. No selected configuration differs significantly from operand-only MX: the product and accumulator precisions change pass@1 by $+3.7$, $-3.0$ and $+0.6$ points for MXFP8, MXFP6 and MXFP4, against a paired 95\% interval of about $\pm$5.6--6.7 points. MXFP8 and MXFP6 are no worse than all-BF16. MXFP4 loses 5.5 points, and that loss comes from operand quantization (operand-only MXFP4 loses 6.1), not from the selected product and accumulators. The $+6.1$ for operand-only MXFP6 is within the spread expected over these comparisons and is not an improvement we claim.

\begin{table}[H]
\centering\small
\setlength{\tabcolsep}{4pt}
\begin{tabular}{lccc}
\toprule
Configuration & pass@1 (\%) & $\Delta$ vs BF16 & $\Delta$ vs op.-only \\
\midrule
All-BF16            & 33.5 &  &  \\
Op.-only MXFP8      & 34.1 & +0.6 &  \\
Op.-only MXFP6      & 39.6 & +6.1 &  \\
Op.-only MXFP4      & 27.4 & $-$6.1 &  \\
Selected MXFP8      & 37.8 & +4.3 & +3.7 \\
Selected MXFP6      & 36.6 & +3.0 & $-$3.0 \\
Selected MXFP4      & 28.0 & $-$5.5 & +0.6 \\
\bottomrule
\end{tabular}
\caption{\textbf{Llama-3.1-8B HumanEval pass@1 at $16{\times}16$.} 164 problems, 0-shot, greedy. $\Delta$ in percentage points, paired over the same problems (95\% interval about $\pm$5--7 points).}
\label{tab:humaneval}
\end{table}

\begin{table*}[!p]
\centering
\begingroup
\small
\setlength{\tabcolsep}{3pt}
\renewcommand{\arraystretch}{0.87}
\setlength{\abovecaptionskip}{3pt}
\setlength{\belowcaptionskip}{0pt}
\resizebox{\textwidth}{!}{%
\begin{tabular}{llccccccccccc}
\toprule
Format & Size & Perplexity & \shortstack{$\Delta$ vs\\BF16} & \shortstack{$\Delta$ vs\\op.-only} & \shortstack{Energy\\(pJ/op)} & $\Delta$ & \shortstack{Area\\(mm$^2$)} & $\Delta$ & \shortstack{Power\\(mW)} & $\Delta$ & \shortstack{Cycles\\($10^9$)} & $\Delta$ \\
\midrule
\multirow{3}{*}{MXFP8} & $8{\times}8$ & 5.8425 & +0.68\% & +0.38\% & 6.39 & -23.7\% & 0.115 & -34.5\% & 197.4 & -23.7\% & 4.19 & -6.4\% \\
 & $16{\times}16$ & 5.7580 & +0.54\% & +0.22\% & 4.19 & -30.7\% & 0.471 & -32.4\% & 517.2 & -30.7\% & 1.15 & -48.3\% \\
 & $32{\times}32$ & 5.7262 & +0.68\% & +0.28\% & 3.65 & -32.5\% & 1.892 & -32.0\% & 1804.5 & -32.5\% & 1.00 & -45.9\% \\
\midrule
\multirow{3}{*}{MXFP6} & $8{\times}8$ & 5.9260 & +2.12\% & +0.72\% & 1.65 & -28.6\% & 0.108 & -38.5\% & 203.8$^*$ & -28.6\% & 4.19 & -6.4\% \\
 & $16{\times}16$ & 5.8490 & +2.13\% & +0.66\% & 1.12 & -34.5\% & 0.445 & -36.2\% & 553.4$^*$ & -34.5\% & 1.14 & -48.5\% \\
 & $32{\times}32$ & 5.7973 & +1.93\% & +0.63\% & 0.99 & -36.1\% & 1.820 & -34.6\% & 1960.0 & -36.1\% & 0.61 & -66.9\% \\
\midrule
\multirow{3}{*}{MXFP4} & $8{\times}8$ & 6.3701 & +9.78\% & +0.30\% & 1.45 & -26.0\% & 0.108 & -38.7\% & 178.6 & -26.0\% & 4.19 & -6.4\% \\
 & $16{\times}16$ & 6.3058 & +10.11\% & +0.54\% & 0.91 & -34.1\% & 0.441 & -36.8\% & 449.6$^*$ & -34.1\% & 1.14 & -48.5\% \\
 & $32{\times}32$ & 6.2901 & +10.60\% & +0.86\% & 0.78 & -36.1\% & 1.803 & -35.2\% & 1541.5$^*$ & -36.1\% & 0.61 & -66.9\% \\
\bottomrule
\end{tabular}%
}
\caption{\textbf{Mistral-7B (v0.3).} BF16 baseline ($8/16/32$): PPL 5.8027 / 5.7270 / 5.6875; cycles 4.48 / 2.22 / 1.85$\times 10^9$.}
\label{tab:results-mistral-7b}
\vspace{5pt}
\resizebox{\textwidth}{!}{%
\begin{tabular}{llccccccccccc}
\toprule
Format & Size & Perplexity & \shortstack{$\Delta$ vs\\BF16} & \shortstack{$\Delta$ vs\\op.-only} & \shortstack{Energy\\(pJ/op)} & $\Delta$ & \shortstack{Area\\(mm$^2$)} & $\Delta$ & \shortstack{Power\\(mW)} & $\Delta$ & \shortstack{Cycles\\($10^9$)} & $\Delta$ \\
\midrule
\multirow{3}{*}{MXFP8} & $8{\times}8$ & 6.7969 & +2.02\% & +1.07\% & 6.39 & -23.7\% & 0.115 & -34.5\% & 197.4 & -23.7\% & 4.19 & -6.4\% \\
 & $16{\times}16$ & 6.6256 & +1.69\% & +0.64\% & 4.19 & -30.7\% & 0.471 & -32.4\% & 517.2 & -30.7\% & 1.15 & -48.3\% \\
 & $32{\times}32$ & 6.5830 & +2.27\% & +1.17\% & 3.65 & -32.5\% & 1.892 & -32.0\% & 1804.5 & -32.5\% & 1.00 & -45.9\% \\
\midrule
\multirow{3}{*}{MXFP6} & $8{\times}8$ & 7.0196 & +5.36\% & +1.42\% & 1.65 & -28.6\% & 0.108 & -38.5\% & 203.8$^*$ & -28.6\% & 4.19 & -6.4\% \\
 & $16{\times}16$ & 6.8558 & +5.23\% & +1.46\% & 1.12 & -34.5\% & 0.445 & -36.2\% & 553.4$^*$ & -34.5\% & 1.14 & -48.5\% \\
 & $32{\times}32$ & 6.8147 & +5.87\% & +1.90\% & 0.99 & -36.1\% & 1.820 & -34.6\% & 1960.0 & -36.1\% & 0.61 & -66.9\% \\
\midrule
\multirow{3}{*}{MXFP4} & $8{\times}8$ & 8.5231 & +27.93\% & +0.71\% & 1.45 & -26.0\% & 0.108 & -38.7\% & 178.6 & -26.0\% & 4.19 & -6.4\% \\
 & $16{\times}16$ & 8.3569 & +28.26\% & +1.09\% & 0.91 & -34.1\% & 0.441 & -36.8\% & 449.6$^*$ & -34.1\% & 1.14 & -48.5\% \\
 & $32{\times}32$ & 8.2729 & +28.52\% & +1.74\% & 0.78 & -36.1\% & 1.803 & -35.2\% & 1541.5$^*$ & -36.1\% & 0.61 & -66.9\% \\
\bottomrule
\end{tabular}%
}
\caption{\textbf{Llama-3.1-8B.} BF16 baseline ($8/16/32$): PPL 6.6623 / 6.5154 / 6.4370; cycles 4.48 / 2.22 / 1.85$\times 10^9$.}
\label{tab:results-llama31-8b}
\vspace{5pt}
\resizebox{\textwidth}{!}{%
\begin{tabular}{llccccccccccc}
\toprule
Format & Size & Perplexity & \shortstack{$\Delta$ vs\\BF16} & \shortstack{$\Delta$ vs\\op.-only} & \shortstack{Energy\\(pJ/op)} & $\Delta$ & \shortstack{Area\\(mm$^2$)} & $\Delta$ & \shortstack{Power\\(mW)} & $\Delta$ & \shortstack{Cycles\\($10^9$)} & $\Delta$ \\
\midrule
\multirow{3}{*}{MXFP8} & $8{\times}8$ & 5.2899 & +0.47\% & +0.28\% & 6.39 & -23.7\% & 0.115 & -34.5\% & 197.4 & -23.7\% & 3.94 & -6.4\% \\
 & $16{\times}16$ & 5.1773 & +0.58\% & +0.28\% & 4.19 & -30.7\% & 0.471 & -32.4\% & 517.2 & -30.7\% & 1.10 & -48.9\% \\
 & $32{\times}32$ & 5.1261 & +0.60\% & +0.15\% & 3.65 & -32.5\% & 1.892 & -32.0\% & 1804.5 & -32.5\% & 0.99 & -47.2\% \\
\midrule
\multirow{3}{*}{MXFP6} & $8{\times}8$ & 5.3917 & +2.40\% & +0.97\% & 1.65 & -28.6\% & 0.108 & -38.5\% & 203.8$^*$ & -28.6\% & 3.94 & -6.4\% \\
 & $16{\times}16$ & 5.3163 & +3.28\% & +1.65\% & 1.12 & -34.5\% & 0.445 & -36.2\% & 553.4$^*$ & -34.5\% & 1.10 & -49.1\% \\
 & $32{\times}32$ & 5.2106 & +2.26\% & +0.68\% & 0.99 & -36.1\% & 1.820 & -34.6\% & 1960.0 & -36.1\% & 0.60 & -67.7\% \\
\midrule
\multirow{3}{*}{MXFP4} & $8{\times}8$ & 6.0384 & +14.68\% & +0.00\% & 1.45 & -26.0\% & 0.108 & -38.7\% & 178.6 & -26.0\% & 3.94 & -6.4\% \\
 & $16{\times}16$ & 5.9343 & +15.29\% & +0.45\% & 0.91 & -34.1\% & 0.441 & -36.8\% & 449.6$^*$ & -34.1\% & 1.10 & -49.1\% \\
 & $32{\times}32$ & 5.8293 & +14.41\% & +0.34\% & 0.78 & -36.1\% & 1.803 & -35.2\% & 1541.5$^*$ & -36.1\% & 0.60 & -67.8\% \\
\bottomrule
\end{tabular}%
}
\caption{\textbf{Llama-2-7B.} BF16 baseline ($8/16/32$): PPL 5.2652 / 5.1473 / 5.0953; cycles 4.21 / 2.16 / 1.87$\times 10^9$.}
\label{tab:results-llama2-7b}
\endgroup
\end{table*}

\begin{table*}[!p]
\centering
\begingroup
\small
\setlength{\tabcolsep}{3pt}
\renewcommand{\arraystretch}{0.87}
\setlength{\abovecaptionskip}{3pt}
\setlength{\belowcaptionskip}{0pt}
\resizebox{\textwidth}{!}{%
\begin{tabular}{llccccccccccc}
\toprule
Format & Size & Perplexity & \shortstack{$\Delta$ vs\\BF16} & \shortstack{$\Delta$ vs\\op.-only} & \shortstack{Energy\\(pJ/op)} & $\Delta$ & \shortstack{Area\\(mm$^2$)} & $\Delta$ & \shortstack{Power\\(mW)} & $\Delta$ & \shortstack{Cycles\\($10^9$)} & $\Delta$ \\
\midrule
\multirow{3}{*}{MXFP8} & $8{\times}8$ & 4.7631 & +0.58\% & +0.47\% & 6.39 & -23.7\% & 0.115 & -34.5\% & 197.4 & -23.7\% & 7.42 & -6.5\% \\
 & $16{\times}16$ & 4.6824 & +0.46\% & +0.24\% & 4.19 & -30.7\% & 0.471 & -32.4\% & 517.2 & -30.7\% & 2.05 & -50.9\%
\\
 & $32{\times}32$ & 4.6431 & +0.56\% & +0.25\% & 3.65 & -32.5\% & 1.892 & -32.0\% & 1804.5 & -32.5\% & 1.87 & -48.5\%
\\
\midrule
\multirow{3}{*}{MXFP6} & $8{\times}8$ & 4.8032 & +1.43\% & +0.35\% & 1.65 & -28.6\% & 0.108 & -38.5\% & 203.8$^*$ & -28.6\% & 7.42 & -6.5\% \\
 & $16{\times}16$ & 4.7388 & +1.67\% & +0.60\% & 1.12 & -34.5\% & 0.445 & -36.2\% & 553.4$^*$ & -34.5\% & 2.04 & -51.1\% \\
 & $32{\times}32$ & 4.7041 & +1.89\% & +0.69\% & 0.99 & -36.1\% & 1.820 & -34.6\% & 1960.0 & -36.1\% & 1.12 & -69.3\% \\
\midrule
\multirow{3}{*}{MXFP4} & $8{\times}8$ & 5.1263 & +8.25\% & +0.23\% & 1.45 & -26.0\% & 0.108 & -38.7\% & 178.6 &
-26.0\% & 7.42 & -6.5\% \\
 & $16{\times}16$ & 5.0219 & +7.74\% & +0.80\% & 0.91 & -34.1\% & 0.441 & -36.8\% & 449.6$^*$ & -34.1\% & 2.04 &
-51.1\% \\
 & $32{\times}32$ & 4.9611 & +7.45\% & +0.14\% & 0.78 & -36.1\% & 1.803 & -35.2\% & 1541.5$^*$ & -36.1\% & 1.11 &
-69.4\% \\
\bottomrule
\end{tabular}%
}
\caption{\textbf{Llama-2-13B.} BF16 baseline ($8/16/32$): PPL 4.7357 / 4.6611 / 4.6170; cycles 7.94 / 4.17 / 3.64$\times 10^9$.}
\label{tab:results-llama2-13b}
\vspace{4pt}
\resizebox{\textwidth}{!}{%
\begin{tabular}{llccccccccccc}
\toprule
Format & Size & Perplexity & \shortstack{$\Delta$ vs\\BF16} & \shortstack{$\Delta$ vs\\op.-only} & \shortstack{Energy\\(pJ/op)} & $\Delta$ & \shortstack{Area\\(mm$^2$)} & $\Delta$ & \shortstack{Power\\(mW)} &$\Delta$ & \shortstack{Cycles\\($10^9$)} & $\Delta$ \\
\midrule
\multirow{3}{*}{MXFP8} & $8{\times}8$ & 5.7727 & +0.95\% & -2.40\% & 6.39 & -23.7\% & 0.115 & -34.5\% & 197.4 & -23.7\% & 7.88 & -6.3\% \\
 & $16{\times}16$ & 5.4376 & +1.75\% & +0.56\% & 4.19 & -30.7\% & 0.471 & -32.4\% & 517.2 & -30.7\% & 2.15 & -50.8\% \\
 & $32{\times}32$ & 5.3753 & +2.87\% & +1.55\% & 3.65 & -32.5\% & 1.892 & -32.0\% & 1804.5 & -32.5\% & 1.95 & -48.9\% \\
\midrule
\multirow{3}{*}{MXFP6} & $8{\times}8$ & 5.9160 & +3.45\% & +0.39\% & 1.65 & -28.6\% & 0.108 & -38.5\% & 203.8$^*$ &-28.6\% & 7.87 & -6.4\% \\
 & $16{\times}16$ & 5.6387 & +5.52\% & +1.14\% & 1.12 & -34.5\% & 0.445 & -36.2\% & 553.4$^*$ & -34.5\% & 2.15 &-51.0\% \\
 & $32{\times}32$ & 5.5432 & +6.09\% & +1.88\% & 0.99 & -36.1\% & 1.820 & -34.6\% & 1960.0 & -36.1\% & 1.15 & -69.7\% \\
\midrule
\multirow{3}{*}{MXFP4} & $8{\times}8$ & 7.1379 & +24.82\% & +1.24\% & 1.45 & -26.0\% & 0.108 & -38.7\% & 178.6 & -26.0\% & 7.87 & -6.4\% \\
 & $16{\times}16$ & 6.8849 & +28.84\% & +0.46\% & 0.91 & -34.1\% & 0.441 & -36.8\% & 449.6$^*$ & -34.1\% & 2.15 & -51.0\% \\
 & $32{\times}32$ & 6.7903 & +29.95\% & +0.76\% & 0.78 & -36.1\% & 1.803 & -35.2\% & 1541.5$^*$ & -36.1\% & 1.15 & -69.7\% \\
\bottomrule
\end{tabular}%
}
\caption{\textbf{Qwen2.5-14B.} BF16 baseline ($8/16/32$): PPL 5.7186 / 5.3438 / 5.2251; cycles 8.41 / 4.38 /3.81$\times 10^9$.}
\label{tab:results-qwen25-14b}
\vspace{4pt}
\resizebox{\textwidth}{!}{%
\begin{tabular}{llccccccccc}
\toprule
Size & Configuration & ARC-e & ARC-c & HellaSwag & PIQA & WinoGrande & BoolQ & LAMBADA & Mean & $\Delta$ mean \\
\midrule
\multirow{7}{*}{$8{\times}8$} & All-BF16 & 54.0 & 31.8 & 56.8 & 76.0 & 58.0 & 62.2 & 59.6 & 56.91 &  \\
 & Op.-only MXFP8 & 55.0 {\scriptsize(+1.0)} & 31.8 {\scriptsize(+0.0)} & 56.2 {\scriptsize($-$0.6)} & 75.4{\scriptsize($-$0.6)} & 57.8 {\scriptsize($-$0.2)} & 61.8 {\scriptsize($-$0.4)} & 59.4 {\scriptsize($-$0.2)} & 56.77 & $-$0.14 \\
 & Op.-only MXFP6 & 55.8 {\scriptsize(+1.8)} & 30.0 {\scriptsize($-$1.8)} & 56.4 {\scriptsize($-$0.4)} & 74.8 {\scriptsize($-$1.2)} & 58.2 {\scriptsize(+0.2)} & 63.0 {\scriptsize(+0.8)} & 59.0 {\scriptsize($-$0.6)} & 56.74 &$-$0.17 \\
 & Op.-only MXFP4 & 50.8 {\scriptsize($-$3.2)} & 31.8 {\scriptsize(+0.0)} & 54.0 {\scriptsize($-$2.8)} & 72.0{\scriptsize($-$4.0)} & 55.6 {\scriptsize($-$2.4)} & 61.2 {\scriptsize($-$1.0)} & 53.2 {\scriptsize($-$6.4)} & 54.09 & $-$2.83 \\
 & Selected MXFP8 & 54.4 {\scriptsize(+0.4)} & 31.4 {\scriptsize($-$0.4)} & 56.0 {\scriptsize($-$0.8)} & 75.4 {\scriptsize($-$0.6)} & 58.2 {\scriptsize(+0.2)} & 62.0 {\scriptsize($-$0.2)} & 59.6 {\scriptsize(+0.0)} & 56.71 &$-$0.20 \\
 & Selected MXFP6 & 55.2 {\scriptsize(+1.2)} & 32.2 {\scriptsize(+0.4)} & 56.0 {\scriptsize($-$0.8)} & 72.8{\scriptsize($-$3.2)} & 58.0 {\scriptsize(+0.0)} & 61.8 {\scriptsize($-$0.4)} & 57.4 {\scriptsize($-$2.2)} & 56.20 & $-$0.71 \\
 & Selected MXFP4 & 51.2 {\scriptsize($-$2.8)} & 31.8 {\scriptsize(+0.0)} & 54.6 {\scriptsize($-$2.2)} & 72.4 {\scriptsize($-$3.6)} & 57.4 {\scriptsize($-$0.6)} & 61.6 {\scriptsize($-$0.6)} & 53.2 {\scriptsize($-$6.4)} & 54.60& $-$2.31 \\
\midrule
\multirow{7}{*}{$16{\times}16$} & All-BF16 & 55.0 & 32.2 & 56.4 & 76.6 & 57.4 & 61.0 & 58.6 & 56.74 &  \\
 & Op.-only MXFP8 & 54.6 {\scriptsize($-$0.4)} & 32.8 {\scriptsize(+0.6)} & 57.2 {\scriptsize(+0.8)} & 75.2{\scriptsize($-$1.4)} & 59.2 {\scriptsize(+1.8)} & 62.2 {\scriptsize(+1.2)} & 59.4 {\scriptsize(+0.8)} & 57.23 & +0.49 \\
 & Op.-only MXFP6 & 53.0 {\scriptsize($-$2.0)} & 31.2 {\scriptsize($-$1.0)} & 57.0 {\scriptsize(+0.6)} & 74.2 {\scriptsize($-$2.4)} & 57.2 {\scriptsize($-$0.2)} & 62.6 {\scriptsize(+1.6)} & 58.0 {\scriptsize($-$0.6)} & 56.17 &$-$0.57 \\
 & Op.-only MXFP4 & 51.8 {\scriptsize($-$3.2)} & 32.2 {\scriptsize(+0.0)} & 53.4 {\scriptsize($-$3.0)} & 73.2{\scriptsize($-$3.4)} & 58.0 {\scriptsize(+0.6)} & 63.8 {\scriptsize(+2.8)} & 54.4 {\scriptsize($-$4.2)} & 55.26 & $-$1.49 \\
 & Selected MXFP8 & 54.4 {\scriptsize($-$0.6)} & 32.0 {\scriptsize($-$0.2)} & 57.6 {\scriptsize(+1.2)} & 75.6 {\scriptsize($-$1.0)} & 60.0 {\scriptsize(+2.6)} & 62.8 {\scriptsize(+1.8)} & 60.0 {\scriptsize(+1.4)} & 57.49 &+0.74 \\
 & Selected MXFP6 & 54.8 {\scriptsize($-$0.2)} & 31.2 {\scriptsize($-$1.0)} & 55.8 {\scriptsize($-$0.6)} & 74.0{\scriptsize($-$2.6)} & 59.4 {\scriptsize(+2.0)} & 63.2 {\scriptsize(+2.2)} & 58.8 {\scriptsize(+0.2)} & 56.74 & +0.00 \\
 & Selected MXFP4 & 54.2 {\scriptsize($-$0.8)} & 32.2 {\scriptsize(+0.0)} & 54.6 {\scriptsize($-$1.8)} & 73.2 {\scriptsize($-$3.4)} & 54.8 {\scriptsize($-$2.6)} & 61.0 {\scriptsize(+0.0)} & 53.0 {\scriptsize($-$5.6)} & 54.71 &$-$2.03 \\
\midrule
\multirow{7}{*}{$32{\times}32$} & All-BF16 & 54.8 & 33.0 & 56.4 & 76.2 & 58.4 & 61.8 & 59.4 & 57.14 &  \\
 & Op.-only MXFP8 & 54.8 {\scriptsize(+0.0)} & 31.4 {\scriptsize($-$1.6)} & 56.8 {\scriptsize(+0.4)} & 76.0{\scriptsize($-$0.2)} & 60.0 {\scriptsize(+1.6)} & 62.4 {\scriptsize(+0.6)} & 59.2 {\scriptsize($-$0.2)} & 57.23 & +0.09 \\
 & Op.-only MXFP6 & 54.0 {\scriptsize($-$0.8)} & 31.6 {\scriptsize($-$1.4)} & 54.8 {\scriptsize($-$1.6)} & 74.2 {\scriptsize($-$2.0)} & 60.4 {\scriptsize(+2.0)} & 61.4 {\scriptsize($-$0.4)} & 58.6 {\scriptsize($-$0.8)} & 56.43 &$-$0.71 \\
 & Op.-only MXFP4 & 52.4 {\scriptsize($-$2.4)} & 30.6 {\scriptsize($-$2.4)} & 53.8 {\scriptsize($-$2.6)} & 72.8{\scriptsize($-$3.4)} & 55.6 {\scriptsize($-$2.8)} & 61.2 {\scriptsize($-$0.6)} & 51.8 {\scriptsize($-$7.6)} & 54.03 & $-$3.11 \\
 & Selected MXFP8 & 53.8 {\scriptsize($-$1.0)} & 31.2 {\scriptsize($-$1.8)} & 56.4 {\scriptsize(+0.0)} & 76.0 {\scriptsize($-$0.2)} & 61.0 {\scriptsize(+2.6)} & 62.4 {\scriptsize(+0.6)} & 59.0 {\scriptsize($-$0.4)} & 57.11 &$-$0.03 \\
 & Selected MXFP6 & 53.4 {\scriptsize($-$1.4)} & 33.0 {\scriptsize(+0.0)} & 55.2 {\scriptsize($-$1.2)} & 75.0{\scriptsize($-$1.2)} & 56.0 {\scriptsize($-$2.4)} & 62.8 {\scriptsize(+1.0)} & 58.2 {\scriptsize($-$1.2)} & 56.23 & $-$0.91 \\
 & Selected MXFP4 & 51.4 {\scriptsize($-$3.4)} & 32.6 {\scriptsize($-$0.4)} & 53.2 {\scriptsize($-$3.2)} & 74.0 {\scriptsize($-$2.2)} & 57.4 {\scriptsize($-$1.0)} & 61.6 {\scriptsize($-$0.2)} & 52.6 {\scriptsize($-$6.8)} & 54.69& $-$2.46 \\
\bottomrule
\end{tabular}%
}
\caption{\textbf{TinyLlama zero-shot accuracy (\%).} First 500 examples per task (lm-eval); \texttt{acc\_norm} for ARC-e/c, HellaSwag, PIQA; \texttt{acc} for WinoGrande, BoolQ, LAMBADA (OpenAI). Parentheses: percentage-point change from same-size all-BF16; op.-only uses lossless products and BF16 accumulation. Approximate 95\% task uncertainty: $\pm4$ points.}
\label{tab:zeroshot}
\endgroup
\end{table*}

\end{document}